\documentclass[useAMS,usenatbib]{mn2e} 
\usepackage{graphicx}
\usepackage{amsmath} 
\title{Constraining the population of intermediate mass black holes by gamma-ray data of the Fornax cluster} 
\author[Chan]{Man Ho Chan \thanks{chanmh@eduhk.hk}
\\ Department of Science and Environmental Studies, The Education University of Hong Kong, Tai Po, Hong Kong}

\begin{document}

\date{Accepted XXXX, Received XXXX}

\pagerange{\pageref{firstpage}--\pageref{lastpage}} \pubyear{XXXX}

\maketitle

\label{firstpage}

\date{\today}

\begin{abstract}
Recent studies of merging black holes suggest that the cosmological mass fraction of primordial black holes (PBHs) is not completely negligible. The mass of a PBH can be as massive as an intermediate mass black hole (IMBH). In this article, we assume that dark matter particles are thermal relic particles and they can self-annihilate. The dark matter around an IMBH may have formed a density spike so that the rate of dark matter annihilation would be greatly enhanced. Using the gamma-ray data of the Fornax cluster and assuming merging events in the cluster are not important, we show that the amount of IMBHs (including PBHs) is very sub-dominant compared with the amount of baryons and dark matter. The upper limit on the IMBH-to-baryon ratio is $\sim 7\times 10^{-4}$ for $m_{\rm DM} \le 10$ TeV.
\end{abstract}

\begin{keywords}
Dark matter
\end{keywords}

\section{Introduction}
The existence of intermediate mass black holes (IMBHs) ($10^2-10^5M_{\odot}$) has been suggested for a long time. They can be relics of evolved Population III stars \citep{Koliopanos}. Some studies even suggest that the primordial black holes (PBHs) formed at the end of inflation in some inflationary models can be as massive as an IMBH \citep{Carr}. Recent theories propose that the IMBHs can be seeds for supermassive black holes (SMBHs) \citep{Volonteri}. Computer simulations of SMBH formation show that the amount of IMBHs can be as large as 0.1\% of total baryons \citep{Islam,Rashkov}.

On the other hand, the idea that PBHs with mass close to the lower mass regime of an IMBH ($M_{\rm BH} \sim 10^2-10^3M_{\odot}$) could contribute a significant amount of dark matter has been revived after the recent detection of binary merging black holes \citep{Ligo} (for a review see \citet{Carr} and references therein). Some studies claim that PBHs can account for all dark matter \citep{Bird,Clesse}. However, other observations such as gravitational micro-lensing and cosmic microwave background detection do not support this idea \citep{Tisserand,Haimoud}. Recent analyses of PBH merger rate suggest that the amount of PBHs could be just 1\% of dark matter \citep{Sasaki}. In other words, the existence of PBHs cannot completely account for the origin of dark matter so that dark matter particles are still required to explain the missing mass in galaxies and galaxy clusters. 

Recently, \citet{Lacroix} suggested that a centrally concentrated relic population of IMBHs, along with ambient dark matter, could account for the Fermi gamma-ray excess in our Galactic Center (e.g., \citet{Goodenough,Calore,Daylan,Ajello}). However, the required annihilation cross section $\sigma v$ is much smaller than the thermal relic cross section predicted by the benchmark model of dark matter production via the thermal freeze-out mechanism. Also, the assumptions of mass segregation and conservation of angular momentum are required for further justification so that this model can really account for the gamma-ray morphology at the Galactic Center. 

In this article, we follow the idea of \citet{Lacroix} and consider the IMBHs with mass $M_{\rm BH} \sim 10^2-10^3M_{\odot}$. By using the gamma-ray data of the Fornax cluster and assuming dark matter particles can self-annihilate with the thermal relic cross section, we can constrain the population of IMBHs without any assumption of IMBH distribution. We show that the IMBHs with mass $M_{\rm BH} \sim 10^2-10^3M_{\odot}$ contribute less than 0.1\% of total baryons. Here, IMBHs include PBHs and other BHs produced in the early universe such as remnants of Population III stars. Therefore, the constraints for IMBHs also apply to PBHs. We use the data of the Fornax cluster because it has the smallest gamma-ray flux to total baryonic mass ratio among the sample of galaxy clusters in \citet{Ackermann}. The small ratio can give a tight constraint for the IMBH-baryon ratio. Previous studies also show that the gamma-ray data of the Fornax cluster can give stringent constraints for dark matter annihilation \citep{Chan}.

\section{Dark matter annihilation from dark matter spikes around IMBHs}
Let's assume that there is a significant amount of IMBHs formed during the Big Bang or the Population III phase. Some of them may have accreted matter or merged together to form SMBHs. Apart from these cases, most of them are distributed throughout galaxies and galaxy clusters. If dark matter particles also exist, their distribution would be influenced by those IMBHs nearby \citep{Gondolo}. The benchmark model reveals that the dark matter profile would form a mini-spike \citep{Zhao,Bertone2,Bertone3,Lacroix}, which can be described as follows:
\begin{equation}
\rho(r)= \left \{ \begin{array}{lll}
0       & {\ \ r \le 2R_S } \\ &\\
\rho_{\rm sat}       & {\ \ 2R_S < r \le R_{\rm sat} } \\ &\\
\rho_0 \left(\frac{r}{R_{\rm sp}} \right)^{-\gamma_{\rm sp}},       & {\ \ R_{\rm sat} < r \le R_{\rm sp} } \end{array} \right.
\end{equation} 
where $\rho_{\rm sat}=m_{\rm DM}/(\sigma vt_{\rm BH})$ is the saturation density with $t_{\rm BH}$ is the age of the IMBH, and $R_{\rm sat}=R_{\rm sp}(\rho_{\rm sat}/\rho_0)^{-1/\gamma_{\rm sp}}$. Here, the normalization density $\rho_0$ can be determined by assuming the mass inside the spike $M_{\rm sp} \approx M_{\rm BH}$ \citep{Lacroix}. This gives $\rho_0 \approx (3-\gamma_{\rm sp})M_{\rm BH}/(4\pi R_{\rm sp}^3)$. The spike slope is $\gamma_{\rm sp}=9/4$ for an adiabatic mini-spike and $\gamma_{\rm sp}=3/2$ for dynamically heated spike \citep{Gnedin}. Therefore, we can assume that $3/2 \le \gamma_{\rm sp} \le 9/4$. Note that the value of $\gamma_{\rm sp}$ would decrease to $\approx 0.5$ due to mergers of IMBHs \citep{Merritt}. We will discuss this special case later.

The total integrated gamma-ray emission rate (in ph s$^{-1}$) for one mini-spike is
\begin{equation}
\Phi_{\rm sp}= \frac{\sigma v}{2m_{\rm DM}^2} \int_0^{R_{\rm sp}} 4\pi r^2 \rho^2(r)dr \int \frac{dN}{dE_{\gamma}}dE_{\gamma},
\end{equation}
where $dN/dE_{\gamma}$ is the photon spectrum per dark matter annihilation, which can be obtained in \citet{Cirelli}. 

Let us assume that the typical IMBH mass is $M_{\rm BH}=10^2M_{\odot}-10^3M_{\odot}$. By extrapolating the relation of supermassive black hole mass and velocity dispersion ($M_{\rm BH}-\sigma_*$ relation) \citep{Tremaine}, we have $R_{\rm sp} \approx GM_{\rm BH}/\sigma_*^2=0.043$ pc for $M_{\rm BH}=10^3M_{\odot}$ and $R_{\rm sp}=0.012$ pc for $M_{\rm BH}=10^2M_{\odot}$. In standard cosmology, the simplest model suggests that dark matter particles were thermally produced after the Big Bang. The annihilation cross section for thermal relic dark matter particles is $\sigma v=2.2\times 10^{-26}$ cm$^3$ s$^{-1}$ \citep{Steigman}. For this cross section, $m_{\rm DM} \ge 100$ GeV is still compatible with the radio \citep{Chan2,Chan3}, antiproton \citep{Cavasonza} and gamma-ray constraints \citep{Ackermann2,Chan}. Although some studies show considerable tensions for certain annihilation channels \citep{Giesen,Chang}, a more recent analysis combining gamma-ray, antiproton and cosmic microwave background data generally allows $m_{\rm DM} \ge 100$ GeV for the thermal relic cross section \citep{Leane}. By combining Eqs.~(1) and (2) with $t_{\rm BH}=10^{10}$ yr and taking the thermal relic cross section, the photon rate becomes \citep{Lacroix}
\begin{equation}
\begin{aligned}
\Phi_{\rm sp}
=& 1.4 \times 10^{39} m_{100}^{-4/3} M_{\rm BH,3}^{4/3} \left( \frac{R_{\rm sp}}{0.043~\rm pc} \right)^{-1} \left( \frac{\sigma v}{2.2 \times 10^{-26}~\rm cm^3~s^{-1}} \right)^{1/3} 
\\
& \times \int \frac{dN}{dE_{\gamma}}dE_{\gamma}~{\rm ph~s^{-1}}
\end{aligned}
\end{equation}
for $\gamma_{\rm sp}=9/4$ and 
\begin{equation}
\begin{aligned}
\Phi_{\rm sp}
=& 6.8 \times 10^{37}m_{100}^{-2}M_{\rm BH,3}^2 \left( \frac{\sigma v}{2.2 \times 10^{-26}~\rm cm^3~s^{-1}} \right) \left(\frac{R_{\rm sp}}{0.043~\rm pc} \right)^{-3}
\\
& \times \ln \left[273M_{\rm BH,3}^{-1}m_{100} \left( \frac{\sigma v}{2.2 \times 10^{-26}~\rm cm^3~s^{-1}} \right)^{-1} \left( \frac{R_{\rm sp}}{0.043~\rm pc} \right)^3 \right] 
\\
& \times \int \frac{dN}{dE_{\gamma}}dE_{\gamma}~{\rm ph~s^{-1}}
\end{aligned}
\end{equation}
for $\gamma_{\rm sp}=3/2$, where $M_{\rm BH,3}=M_{\rm BH}/10^3M_{\odot}$ and $m_{100}=m_{\rm DM}/100~\rm GeV$.

We use the gamma-ray data of the Fornax cluster to constrain the amount of IMBHs. The upper limits (point-source) of the total gamma-ray flux $\phi$ for $E_{\gamma}>0.5$ GeV, $E_{\gamma}>1$ GeV and $E_{\gamma}>10$ GeV are $F_{\gamma}=1.1 \times 10^{-10}$ ph cm$^{-2}$ s$^{-1}$, $F_{\gamma}=4.9 \times 10^{-11}$ ph cm$^{-2}$ s$^{-1}$ and $F_{\gamma}=2.6 \times 10^{-12}$ ph cm$^{-2}$ s$^{-1}$ respectively \citep{Ackermann}. In particular, the data for $E_{\gamma}>10$ GeV can give the most stringent limits as the ratio $\Phi/N_{\gamma}$ is the smallest (see Fig.~1), where $\Phi$ is the total photon emission rate and $N_{\gamma}=\int (dN/dE_{\gamma})dE_{\gamma}$ (the upper limit of $E_{\gamma}$ is 200 GeV). By taking the distance to the cluster $D=19.8$ Mpc \citep{Sanchez} and using the point-source approximation, the total photon emission rate is $\Phi=4\pi D^2 \phi=1.23 \times 10^{41}$ ph s$^{-1}$. If we saturate the gamma-ray upper limit with dark matter annihilation due to mini-spikes $\Phi=N_{\rm BH}\Phi_{\rm sp}$, we can get an upper limit on the amount of IMBHs $N_{\rm BH}$. Note that we have not considered the annihilation flux due to the smooth dark matter halo. However, accounting for this would lead to more stringent constraints, so the limit obtained in this analysis is more conservative.

The total mass of hot gas in the Fornax cluster is $M_{\rm gas}=0.21^{+0.05}_{-0.06} \times 10^{13}M_{\odot}$ (assuming $h=0.68$ for the Hubble constant) \citep{Chen}. Since the baryonic mass is dominated by the hot gas in a galaxy cluster, we can write the IMBH-to-baryon ratio $f$ as
\begin{equation}
f \approx \frac{N_{\rm BH}M_{\rm BH}}{M_{\rm gas}}= \frac{\Phi M_{\rm BH}}{\Phi_{\rm sp}M_{\rm gas}}.
\end{equation}
By considering 4 popular annihilation channels ($e^+e^-$, $\mu^+\mu^-$, $\tau^+\tau^-$ and $b\bar{b}$) and 2 limits of $\gamma_{\rm sp}$ (i.e. $\gamma_{\rm sp}=9/4$ and $\gamma_{\rm sp}=3/2$), we can get $f$ as a function of $m_{\rm DM}$ for $M_{\rm BH}=10^2M_{\odot}$ and $M_{\rm BH}=10^3M_{\odot}$. Fig.~2 shows that $f$ is much smaller than $10^{-3}$ for $m_{\rm DM}=100-10000$ GeV. The largest value of $f$ is $\sim 7\times 10^{-4}$ (the $e^+e^-$ and $\mu^+\mu^-$ channels) when $m_{\rm DM}=10000$ GeV. In other words, the total mass of IMBHs in our universe is not significant. This does not satisfy the criterion $\Omega_{\rm IMBH} \sim 10^{-3}\Omega_{\rm baryon}$ derived in \citet{Rashkov} for IMBHs to be the seeds of SMBHs, and challenges the claim that IMBHs could be 1\% of dark matter \citep{Sasaki}, if dark matter particles can indeed self-annihilate and they are the thermal relics.

We can also use the extended flux limits $\phi_{\rm extended}$ (in ph cm$^{-2}$ s$^{-1}$) to constrain the value of $N_{\rm BH}$. By assuming the IMBHs are distributed uniformly throughout the cluster, we have
\begin{equation}
N_{\rm BH}=\frac{\phi_{\rm extended}V}{\Phi_{\rm sp}J},
\end{equation}
where $V$ is the total volume of the Fornax cluster (within the virial radius) and $J$ is the effective `J-factor' which is given by
\begin{equation}
J=\frac{1}{4\pi} \int d\Omega \int dl.
\end{equation}
Here, $\Omega$ is the solid angle and $l$ is the line-of-sight distance. Note that the integral in the above equation only includes the region inside the cluster. Taking the virial radius to be $0.71^{+0.07}_{-0.12}$ Mpc \citep{Chen}, we get $V=1.5^{+0.5}_{-0.6}$ Mpc$^3$ and $J=4.5^{+1.5}_{-2.0}\times 10^{-4}$ Mpc. The ratio of the observed extended flux to point-source flux is $\phi_{\rm extended}/\phi \approx 3$ \citep{Ackermann}, which is comparable to the ratio $4\pi D^2J/V=1.5^{+1.8}_{-0.9}$. However, since we don't know the actual distribution of IMBHs, using the extended limits for calculation might suffer from large systematic uncertainty. Therefore, we use the point-source approximation to do the analysis, which can avoid any unjustified assumption of IMBH distribution.

The above analysis assumed that all the IMBHs have retained mini-spikes of annihilating dark matter. However, IMBH mergers would significantly soften the dark matter density cusps around the IMBHs \citep{Merritt}. The inner slope would be decreased to $\gamma_{\rm sp} \sim 0.5$ if mergers are important \citep{Merritt}. We consider an extreme case in which all IMBHs underwent mergers, and subsequently all the associated mini-spikes have $\gamma_{\rm sp}=0.5$. By putting $\gamma_{\rm sp}=0.5$ in Eq.~(1) and using Eq.~(2), we can calculate the corresponding $\Phi_{\rm sp}$ for this extreme case. Fig.~3 shows that the IMBH-to-baryon fraction can be as large as 1 for $m_{\rm DM} \sim 10$ TeV in this very extreme case. Since the baryon-to-dark matter fraction is about 0.17 \citep{Planck}, the IMBHs can contribute 20\% of dark matter for $m_{\rm DM}$ close to 10 TeV. In this case the requirement for IMBHs to be the seeds of SMBHs ($f \sim 10^{-3}$) would be satisfied if $m_{\rm DM}>1$ TeV. Nevertheless, merging of IMBHs might be somewhat frequent in galaxies, but not in the intergalactic regions of galaxy clusters. The average mass density of baryons in the intergalactic region of a typical galaxy cluster is $\le 10^{-26}$ g cm$^{-3}$ \citep{Chen}. Therefore, the number density of IMBHs in the intergalactic regions of a galaxy cluster should also be as low as this equivalent value ($\le 10^{-6}$ pc$^{-3}$), assuming $M_{\rm BH}=10^2M_{\odot}$. Hence the probability of merging would not be high and this very extreme case is unlikely to happen.

On the other hand, if we allow the annihilation cross section to be a free parameter, we can derive upper limits on the annihilation cross section for $f=10^{-3}$ (assuming $M_{\rm BH}=10^2M_{\odot}$). We can see from Fig.~4 that the resultant upper limits strongly depend on $\gamma_{\rm sp}$. Also, the upper limits are very small, with typically $\sigma v \le 10^{-30}$ cm$^3$ s$^{-1}$ for $m_{\rm DM} \sim 1$ TeV. Such small cross sections may be indicative of velocity-dependent dark matter annihilation.

\begin{figure}
\vskip 10mm
 \includegraphics[width=85mm]{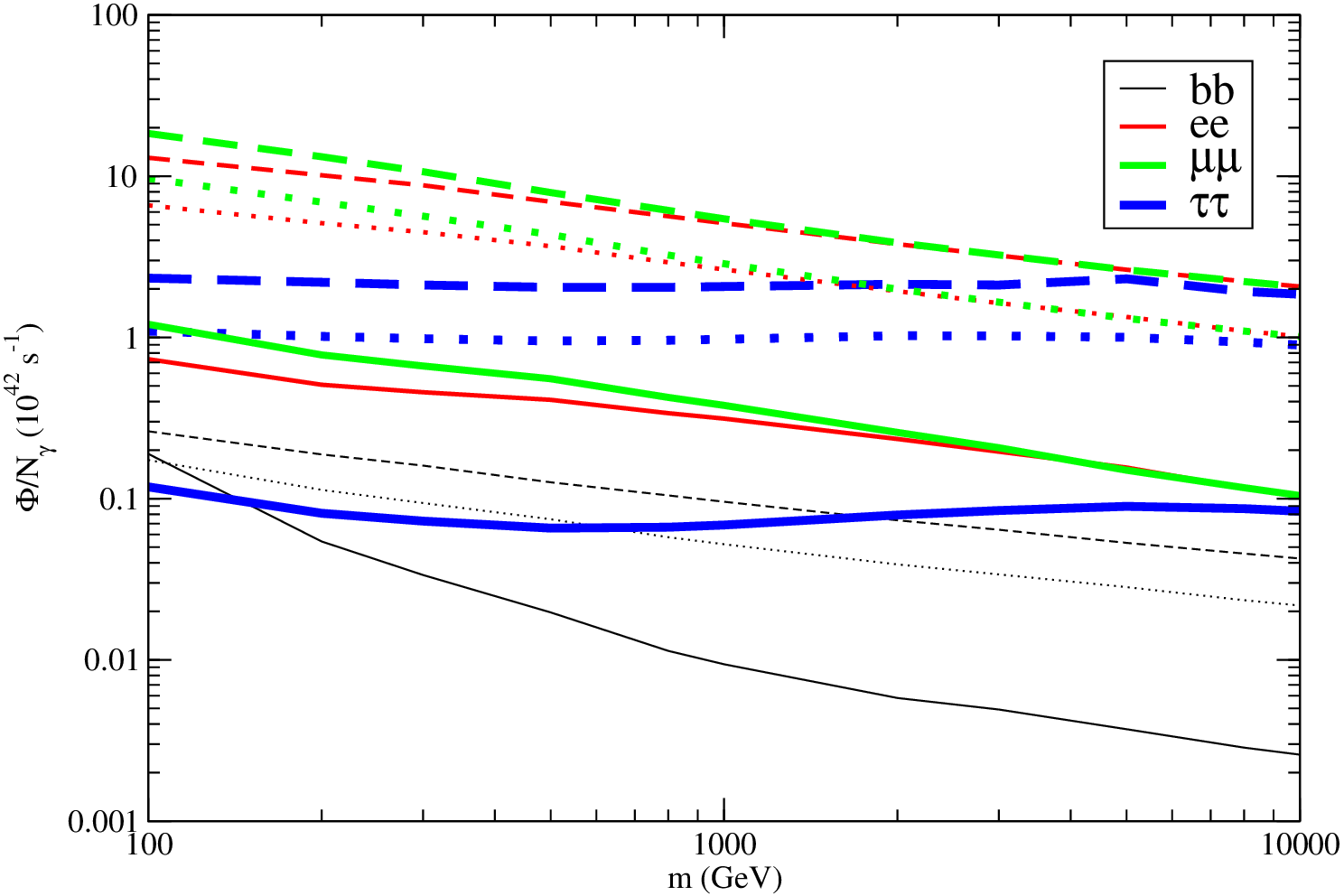}
 \caption{The ratio of $\Phi/N_{\gamma}$. The black, red, green and blue lines represent the corresponding ratios for $b\bar{b}$, $e^+e^-$, $\mu^+\mu^-$ and $\tau^+\tau^-$ channels respectively (solid lines: data for $E_{\gamma}>10$ GeV; dotted lines: data for $E_{\gamma}>1$ GeV; dashed lines: data for $E_{\gamma}>0.5$ GeV.}
\vskip 10mm
\end{figure}

\begin{figure*}
\vskip 10mm
\centering
 \includegraphics[width=150mm]{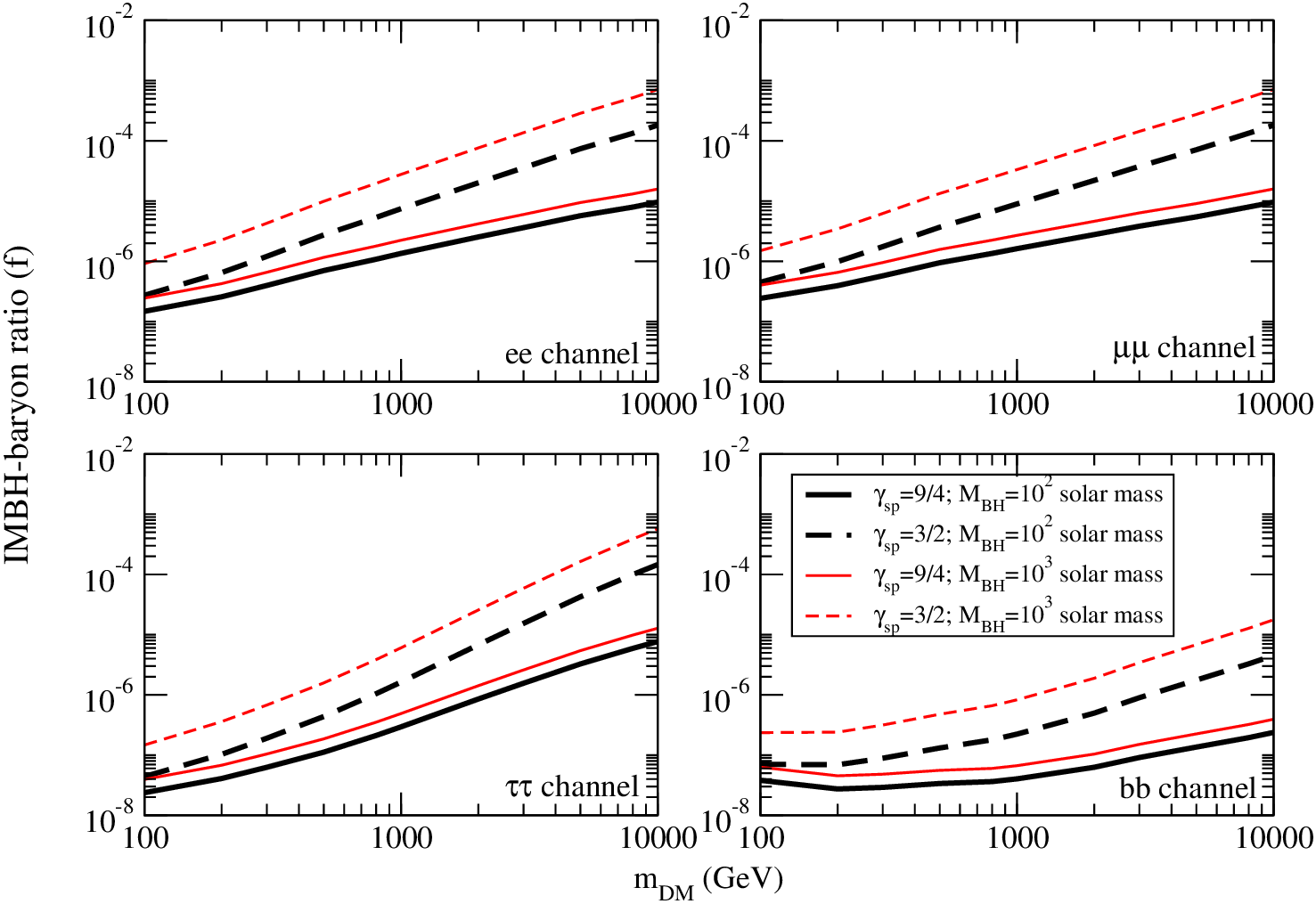}
 \caption{The upper limits of the IMBH-baryon ratio $f$ for 4 popular annihilation channels (black: $M_{\rm BH}=10^2M_{\odot}$; red: $M_{\rm BH}=10^3M_{\odot}$; solid lines: $\gamma_{\rm sp}=9/4$; dashed lines: $\gamma_{\rm sp}=3/2$).}
\vskip 10mm
\end{figure*}

\begin{figure*}
\vskip 10mm
\centering
 \includegraphics[width=150mm]{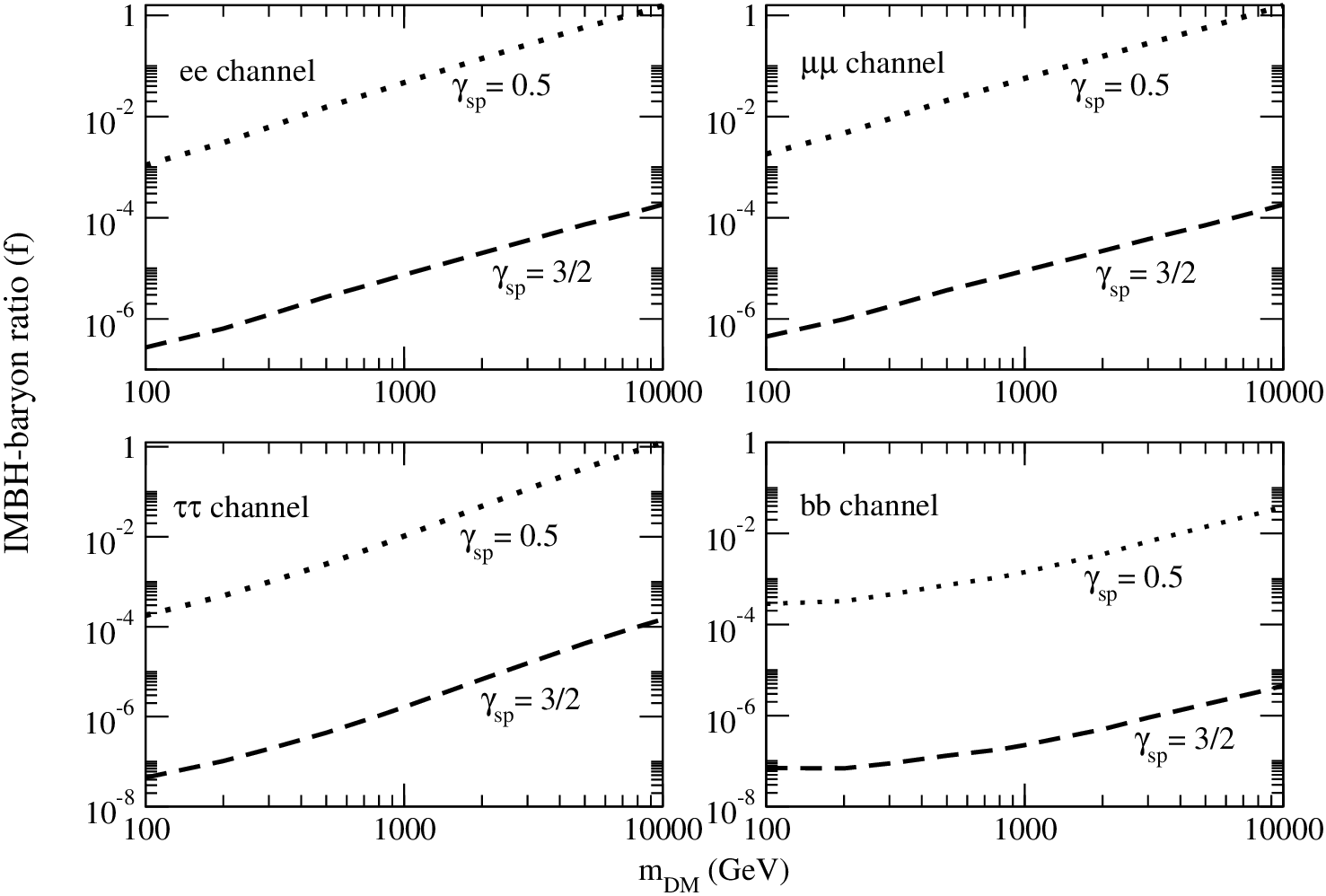}
 \caption{The upper limits of the IMBH-baryon ratio $f$ for 4 popular annihilation channels (dotted lines: $M_{\rm BH}=10^2M_{\odot}$ with $\gamma_{\rm sp}=0.5$; dashed lines: $M_{\rm BH}=10^2M_{\odot}$ with $\gamma_{\rm sp}=3/2$.)}
\vskip 10mm
\end{figure*}

\begin{figure*}
\vskip 10mm
\centering
 \includegraphics[width=150mm]{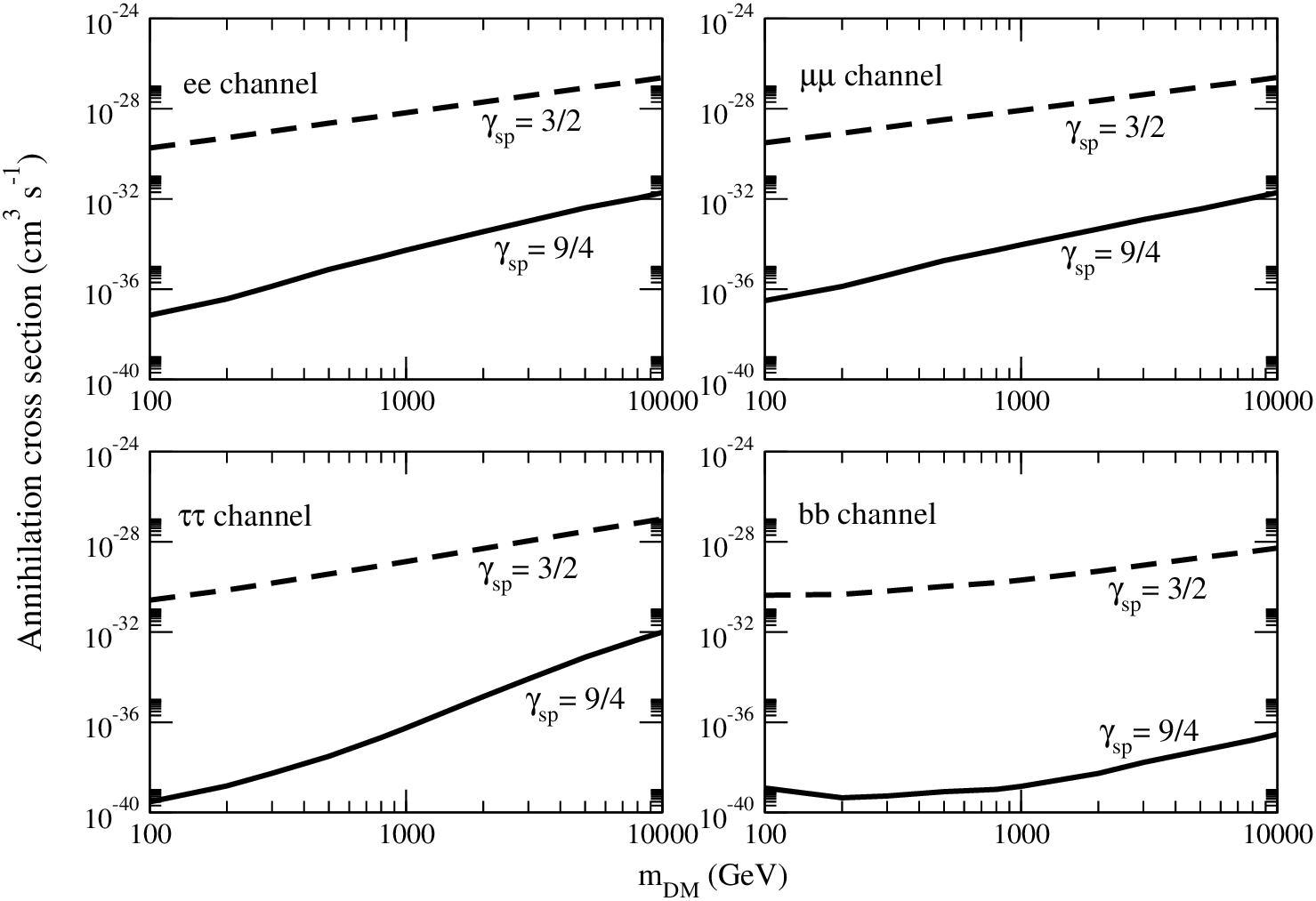}
 \caption{The upper limits of annihilation cross section $\sigma v$ for 4 popular annihilation channels (solid lines: $M_{\rm BH}=10^2M_{\odot}$ with $\gamma_{\rm sp}=9/4$; dashed lines: $M_{\rm BH}=10^2M_{\odot}$ with $\gamma_{\rm sp}=3/2$.)}
\vskip 10mm
\end{figure*}

\section{Discussion}
In this article, by using the gamma-ray data of the Fornax cluster and assuming IMBH mergers are subdominant in the intergalactic medium, we show that the total mass of IMBHs ($M_{\rm BH}=10^2-10^3M_{\odot}$) is not significant, provided that dark matter particles can self-annihilate and they are thermal relics. We find an upper limit on the IMBH-to-baryon ratio of $f \le 7\times 10^{-4}$ for $m_{\rm DM} \le 10$ TeV. Also, $f$ increases with $M_{\rm BH}$. The amount of IMBHs would be much smaller if $M_{\rm BH} \le 10^2M_{\odot}$. For the cosmological benchmark model of weakly interacting massive particle (WIMP) annihilating dark matter suggested in \citet{Bertone}, the dark matter mass is $m_{\rm DM}=218$ GeV with $\sigma v \sim 10^{-26}$ cm$^3$ s$^{-1}$. Our results suggest that $f<10^{-5}$ for this benchmark model, which means that the population of IMBHs is not a significant component in our universe. Therefore, our results challenge the suggestion based on the merging black hole rate that IMBHs could be $\sim 1$\% of dark matter \citep{Sasaki}. However, this suggestion can still endure if $m_{\rm DM} \gg 10$ TeV or the annihilation cross section is much smaller than the thermal relic cross section. It then requires some special scenarios to account for the production of dark matter (e.g. non-thermal production mechanisms). Also, if merging events are very frequent so that most of the density spikes are softened to $\gamma_{\rm sp}=0.5$, then the allowed amount of IMBHs would be much larger.

\section{acknowledgements}
This work is supported by a grant from The Education University of Hong Kong (activity code: 04256).

\label{lastpage}

\end{document}